# Mutual Clustering Coefficient-based Suspicious-link Detection Approach for Online Social Networks

Mudasir Ahmad Wani                Suraiya Jabin
mudassir148001@st.jmi.ac.in       sjabin@jmi.ac.in

**Department of Computer Science**
Jamia Millia Islamia (A Central University),
New Delhi, India

**Abstract**

Online social networks (OSNs) are trendy and rapid information propagation medium on the web where millions of new connections either positive such as acquaintance or negative such as animosity, are being established every day around the world. The negative links (or sometimes we can say harmful connections) are mostly established by fake profiles as they are being created by minds with ill aims. Detecting negative (or suspicious) links within online users can better aid in mitigation of fake profiles from OSNs.

A modified clustering coefficient formula, named as $Mutual\ Clustering\ Coefficient$ represented by $M_{cc}$, is introduced to quantitatively measure the connectivity between the mutual friends of two connected users in a group. In this paper, we present a classification system based on mutual clustering coefficient and profile information of users to detect the suspicious links within the user communities. Profile information helps us to find the similarity between users. Different similarity measures have been employed to calculate the profile similarity between a connected user pair. Experimental results demonstrate that four basic and easily available features such as $work(w), education(e), home\_town(ht)\ and\ current\_city(cc)$ along with $M_{CC}$ play a vital role in designing a successful classification system for the detection of suspicious links.

## 1. Introduction

In the past decade, the connectivity within people has been spreading rapidly with the help of social networking sites. The connection (or link) of a person on OSNS can be either positive such as friendship or negative such as animosity. The negative links are mostly established by fake profiles as they are being created by minds with ill aims such as running spam campaigns [1], casting unfair online votes [2], accessing user personal information [3], etc. In order to fulfill aims, the fake profile users need to create as many links as possible with real profiles. The chances of friend request being accepted by a real user from the fake profiles are low as most of the connections are being established on a network if the two persons either know each other in offline or share some interests. Therefore, in order to increase the chance of friend requests being accepted, the fake users are nowadays targeting the communities[1] where users are connected and share a strong bond with each other. It is very frequent to have a high number of mutual friends for members of a group of connected people. Furthermore, it has been realized that more is the number of mutual friends, the more is the possibility that the friend request is being accepted by the users [31]. This particular feature of common friends is exploited by the fake profile owners to increase their coverage. Once these fake users succeed in bobbing few naïve users in a group (or on a page), their trust level gets increased among other members in the form of common neighbors which increases the acceptance chances of friend requests by others members in the group. Since these fake users are getting penetrated into the online communities in a clever way and get cloaked into the real public, therefore, it becomes challenging for researchers to identify and obliterate them from the social networks.

---
[1] Communities we here mean the pages or groups created by users on Facebook

Research shows that users in OSNs connect with the people either they know in offline or met online. Social networking sites such as Facebook are being primarily used by people to maintain and strengthen the pre-existing offline social relations. It has been observed that if two persons have enough number of common friends, there are high chances that the two persons share some common offline entity such as same organization, school, course, etc., that cause them to befriend online. However, even if somehow the fake users managed to penetrate into the user groups by exploiting the mutual connections, on the other hand, there are least chances of similarity between two profiles. Based on this observation we proposed a novel approach to identify the suspicious links established by the fake users.

In this paper, we present an approach to identify suspicious (negative) links established by the adversaries by exploiting the mutual friend feature in a group or a page on Facebook. Identifying suspicious links can better aid in designing the fake user detection system. The proposed approach is based on the combination of mutual clustering coefficient and profile information of a user which basically assists in detecting suspicious connections in a group or a page on Facebook. Clustering coefficient [22] is one of the topological measures used to study the structure of a graph. For a graph like Facebook-network, the clustering coefficient indicates to which extent people have mutual friends or how likely the friends of a user are connected to each other. High clustering coefficients signify a tightly connected community in which most of the friends of a user are themselves friends. In our work, we have modified the clustering coefficient as Mutual Clustering Coefficient represented by $M_{cc}$, to measure the connectivity between the mutual friends of two connected users in a group. Profile information helps us to find the similarity between users. The similarity between two user profiles based on the selected attribute set can be calculated by several text-based similarity measures such as N-gram [23], Cosine similarity [24], Jaro [19]. Moreover, the authors in [20] have presented more than ten approaches to compare text documents. In this paper Fuzzy string based similarity measures profile similarity between connected friends. The main contributions of the paper are as follows:

- The $M_{cc}$, a novel and a unique feature, discussed in section 3.3 have been introduced first time for the detection of suspicious (negative) links on Facebook.

- The four basic and easily available features including $work(w), education(e), home\_town(ht)$ $and\ current\_city(cc)$ along with $M_{CC}$ have been extracted from users on Facebook network with the help of IMcrawler [43] in order to form the training dataset. The collected data set along with source code has been made available to help researchers of different domains.

- Fuzzy string based similarity measures have been used to efficiently calculate the profile similarity between connected user pairs.

- Fake Identities have been manually designed and injected into the network to establish the links with real users of a community on Facebook network.

- A classification system based on machine learning techniques such as Decision Tree (J48), RBF based Support Vector Machine (SVM) and Naïve Base (NB) has been designed to build proposed suspicious link identification model with 99.60% accuracy.

The proposed model can be used by the OSN service providers to suggest its members with a list of suspicious connections (links) from their respective friend lists so that a user can themselves verify the suggested links and filter their friend list as per their choice. Although the proposed approach has been tested for the Facebook users only, with the little modification it will be applicable to other social network sites as well. The remainder of the paper continues as follows. Section 2 reviews relevant literature on detection of suspicious links and fake profiles in online in OSNs. Section 3 describes four stages of proposed work towards detection of suspicious links. First two stages deal with the collection of data and its preprocessing. Construction of features and their analysis are carried out in the third stage. Model training and validation along with experimental study are conducted in fourth stage.

## 2. Related Work

In OSNs such as Facebook, a friendship connection is a relationship between two users which gets initiated and established when one user sends a friend request and another user accepts the request respectively. However, this is not true, that the users will always have well-wisher connections in their friend community. Negative links, which in general signifies disapproval, disagreement, distrust, deception, or deceive, can also get established within the users which may lead to bad consequences on the network. Several researches have been carried out to distinguish these negative links from normal (positive) connections. For example, a study [45] conducted on Wikipedia network to classify people who will vote in favor of or against a selection for an adminship. Another study [33] is conducted on three different datasets (Epinions, Slashdot and Wikipedia network) to predict a positive link between two users based on the relationship signs with the friends surrounding them.

Since fake profiles also contribute to the negative links, therefore, in our work, we have used the notion of a negative link to predict the connection established by the fake profiles on the Facebook network. We have utilized a link feature along with some basic profile attributes to predict the status (either normal or negative) of a link in a user community as it has been observed that most of the OSN users freely accept and create connections with the other users on the network without much investigation. According to a study [44], 41% of the Facebook users, who were contacted, accepted the connection request from a random person. It has also been observed that there are more chances of a connection being established if there are already common connections between the two [31]. As we are focusing on the detection of suspicious links on OSNs which aids in designing of a better fake profile detection system, therefore, all the studies carried out for the detection of fake profiles in social networking sites are also related to our work. Most of the literature dealing with detection of fake profiles on OSNs is generally focus on user behavior or profile information of a user and make use of machine learning to solve the problem as discussed in paper [28]. A study [30] has proposed fake profiles detection approach based on the user network structure and identified more than 10 million fake accounts along with 700 million links which were established by fake profiles on Webo[2]. In another approach [12] the authors have applied Bayesian classifier and k-means clustering on profile features including gender and location for deception detection on Twitter Network. A study [34] has used different feature categories including Graph-based features (such as clustering coefficient, betweenness centrality, etc.), Neighbor-based features (such as average neighbors' followers, average neighbors' tweets, etc.), Automation-based features such as average (API ratio, API URL, etc.) and Timing-based feature (such as Following Rate) to detect the spammers on Twitter. Researchers are also exploring other dimensions to mitigate the fake identities from social networking websites. The authors in [11] have presented a technique called SybilGuard [11] to protect a social network from Sybil attacks by differentiating the Sybil nodes from trust nodes using the calculated trust-relationship. The technique is based on the ranking of nodes and a node is considered as a high-ranked node if it is within the local community of a trusted node.Similarly, there are several other graph based features such as groups joined by users, number of friend requests accepted (in degree), number of friend requests sent (out-degree), the extent to which a node acts as a bridge between other nodes (betweenness centrality), nearest node to all other nodes in the network (closeness centrality), growth of OSN graph over time, average degree of nodes and number of singleton friends, etc have been used by researchers [5, 13] for the detection of fake profiles on OSNs

Our work is different from existing studies as it focuses more on tactics and strategies (such as exploitation of 'mutual friends' feature) utilized by current fake users and we further presented new machine learning features to more effectively detect the suspicious links on the Facebook.

---
[2] A Chinese microblogging website

# 3. The Proposed System: Suspicious Link Classifier

Earlier it has been seen that the fake users often face difficulties in establishing friendship relation with the real users as the real users do not easily trust the strangers. In order to bypass this obstacle and gain the trust of users, nowadays the attackers have started targeting user communities rather than the individual user. Targeting a community of users increases the chances of friend requests being accepted by the real users because the members of a particular community are mostly friends to each other and once a member accepts the friend request, the probability of accepting the friend request by other members gets increased. Once the fake user penetrates into some user community, they start sending the friend requests randomly to its members to grow their network. As soon as they trick few users to accept the friend request in the community, they start exploiting the mutual-friend feature to spread their network by sending the friend requests to other members of the community. After establishing connections with a large number of users in the community, they start injecting spam into the network or carry out other unlawful activities. Moreover, in order to gain the confidence over benign users, fake identities target the friends of the victim with the notion that more the number of mutual friends with the victim, more likely the victim will accept the request without much investigation.

As per our empirical analysis, we noted that although the attacker may be successful in establishing the connection (suspicious-link) with the victim by exploiting his friends it is highly unlikely that the attacker and victim have the similar profile attributes. As in real scenario, mutual friendship indicates that there are some common features between the two people that made them befriend, which would not be true for a connection between real and fake user despite having a significant number of mutual friends. In this section, we have presented a framework based on mutual clustering coefficient and profile similarity of connected user pairs to detect the links established by adversarial accounts who have exploited the mutual-friends relationship to grow their network within legitimate users. Basically, our aim in this paper is to build a complete classification system to identify the fake connections within the users of a community on Facebook network. We conducted a social experiment on a Facebook page in order to implement and validate our system.

Figure 1 describes the work flow of the proposed system in four main components, namely Data collection, Data preparation, Feature Construction, and Suspicious link classifier. Each component is discussed in the following subsections.

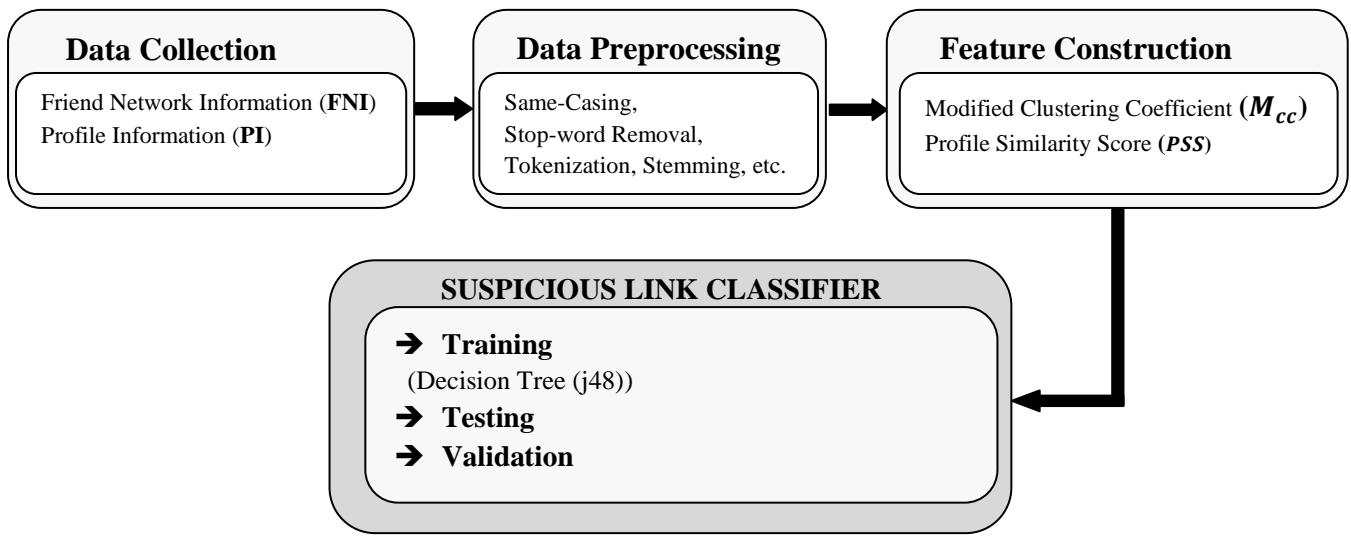

**Figure 1:** Mutual Clustering Coefficient-based suspicious-link identification Framework

## 3.1 Data Collection

In order to build a classifier, collecting dataset is the preliminary step. Since the proposed classifier distinguishes the fake and normal links based on mutual clustering coefficient ($M_{CC}$) of a user pair and their profile similarity, therefore, we need to obtain the information of friend network of the connected user pairs within that particular community and the profile features of both the users as well as their friends have disclosed about themselves. For the collection of required data, we can either use APIs provided by OSN service providers such as Graph API for the Facebook network [15] or by designing own stand-alone crawler program [17, 18]. For the current work, we used IMcrawler [43] in order to extract the data from the user community on the network. IMcrawler is an iMacros-based data crawler, designed to extract every piece of information which is accessible through a browser from the Facebook website. From each user profile in the community, we have extracted four features including *work, education, hometown* and *current city* to calculate the user similarity of two friends using fuzzy logic based string matching technique. Although profile similarity may be calculated on various other aspects of users such as to which political party they belong, what are common groups and pages liked, etc., but most of the time, these aspects are generally not revealed by the users on the Facebook network.

The four basic features which we have included in our study are generally not kept private by the users to their own friend network on Facebook and this set of attributes play an important role in measuring the profile similarity of any two connected users. Facebook users generally establish online communities based on educational institutes or working organizations they belong to, etc. For example, the users of the same school, college, university or same organization create a group in order to remain connected and discuss things around.

Furthermore, people can be friends based on the place which they originally belong to ( *home town* ) or the location they are presently living in ( *current city* ).

We extracted the information from the users and their friends on a Facebook page using the IMcrawler. We logged this information from 839 connections established within 77 users on that page with 10 manually injected fake profiles. In our collected data, we manually labeled a link (edge) as suspicious based on its connection to any of the manually injected fake profiles. The fake profiles have successfully created more than 250 suspicious links within the user community. Based on Exploratory Data Analysis (EDA) [42] we observed that there are three categories of links between the users namely suspicious, normal and fake. Suspicious links are different from real and fake ones, as the real/normal connections are the ones created amongst legitimate users only and the fake connections have malignant users on both the ends as illustrated in Figure 2. The description of the collected dataset is depicted in Table 1. The updated dataset along with source code files is available from https://github.com/Mudasir-IIIT-Bangalore/Mcc-based-Suspicious-Link-Detection/

| Description | Number |
|---|---|
| total # edges (Links) | 839 |
| # normal edges (Links) | 587 |
| # suspicious edges (Links) | 252 |
| total # nodes (users) | 77 |
| # real nodes | 67 |
| # fake nodes | 10 |
| average # connections | 11 |

**Table 1:** Statistics of Data collected from Facebook page

## 3.2 Data Preprocessing

The collected data mainly exists in raw form and may contain missing information. The missing value is the common phenomenon in the collection of social network data since the users have the privilege to hide information from other users or friends and most of the fields while registering to the social networks like Facebook are optional. Here, we are not considering the profiles who's profile information is not available publicly or to the friends of friends.

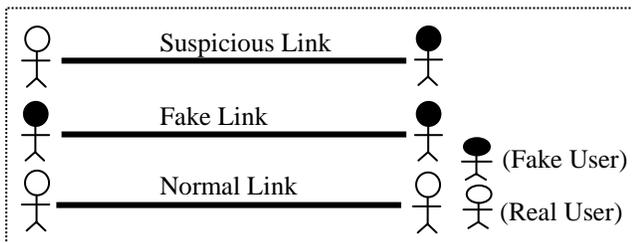

**Figure 2:** Categories of Links in user communities

Before calculating the similarity between users, we applied several text processing techniques including same-casing, stop-word removal, tokenization, and stemming on the extracted user features by using the Natural Language Toolkit (NLTK) [27] library of python programming as shown in Algorithm 2. Stop words such as "the", "a", "an", "in", etc. were removed and the upper and lower case of strings were converted to the same case for the extracted features.

Using the stemming technique, we are converting all the word variations of the same meaning into a root word which makes the overall process of similarity calculation convenient. The actual aim of applying different text analysis techniques here is to prepare the extracted data for applying several similarity measures to calculate the similarity between two connected users. Furthermore, in order to make it more convenient for different similarity measures to calculate the similarity score between two profiles, a word- dictionary of the words has been designed based on the very frequently observed words in the dataset along with their possible word variations. For example, the term *research scholar* has been stored in the dictionary with its word variations like *scholar, researcher, p.hd. scholar*, *ph.d. researcher,* etc. Similarly, other most frequent words have been stored in their corresponding dictionaries.

---

**Algorithm 1** − *TextProcessing*
{
  **INPUT**: $feature\_matrix = extracted\_raw\_features$

  $fm = feature\_matrix[\,][\,];$

  for $(i = 1\ to\ nrows(fm))${
    for $(j = 1\ to\ ncols(fm))${
      $featureText = \textbf{lower}(fm(i,j));$
      $featureText = \textbf{tokenization}(featureText)$
      $featureText = \textbf{stopwords}(featureText);$
      $featureText = \textbf{stemming}(featureText);$
      $norm\_string\ =\ "";$
      for $(i = 1\ to\ length(featureText))$
        $norm\_string\ =\ norm\_string\ +\ featureText(i)\ +\ "\ ";$
    }
  }
}

---

**Figure 3:** Data Normalization using Text Processing techniques

## 3.3 Feature Construction

Feature construction involves designing the input vector from the collected data with the aim to build more optimal features and to design an efficient system. Since the proposed classifier is based on the two aspects, the mutual clustering coefficient and the profile similarity between the two connected users in a community, therefore a feature matrix has been constructed for each pair of connected users in the collected data. The first two columns are reserved for a connected user pair, the third column holds mutual clustering coefficient values derived from friend network information.

Clustering coefficient basically measures the probability to which friends of a user are themselves connected to a network and it can be calculated as follows:
$$CC_v = \frac{2l_v}{k_v(k_v-1)} \qquad (1)$$

other words, we calculate the clustering of two connected users based on their mutual friends only. According to our observation, the strong connection between the mutual friends of a user pair signify that the user pair belongs to a common community and so the users have some sense of similarity in their profiles as well. The modified clustering coefficient between mutual friends of two connected users belonging to same community $U$ can be calculated as follows:

$$Mcc_{(u,v)} = \frac{2l_{u,v}}{m_{u,v}(m_{u,v}-1)}, \forall\, u,v\,\varepsilon\, U \qquad (2)$$

Here, $Mcc_{(u,v)}$ represents the mutual clustering coefficient of user $u$ and $v$, $m_{u,v}$ is the number of mutual friends between $u$ and $v$, and $l_{u,v}$ is the

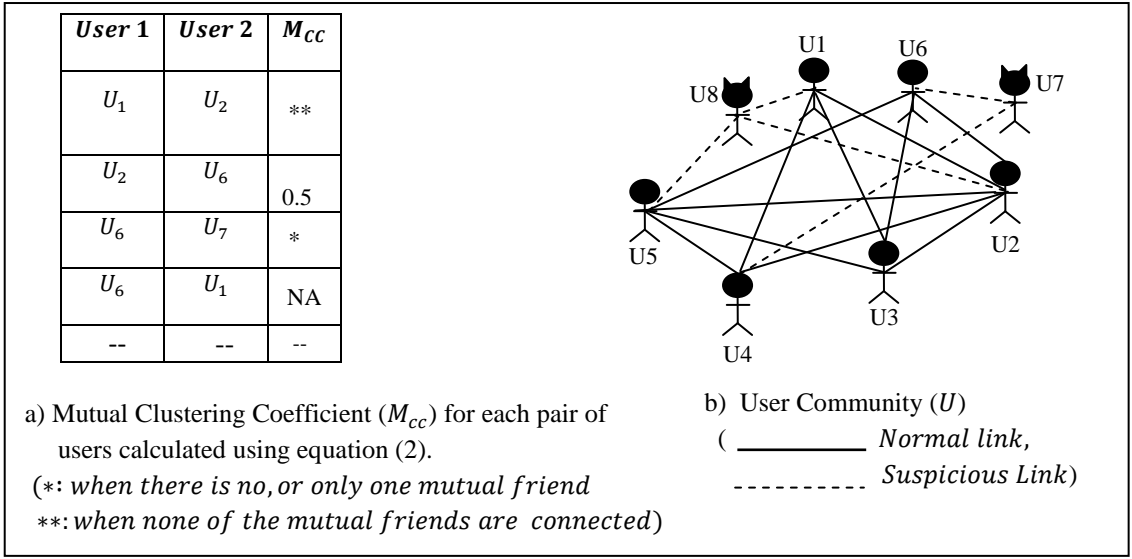

a) Mutual Clustering Coefficient ($M_{cc}$) for each pair of users calculated using equation (2).
(*: when there is no, or only one mutual friend
**: when none of the mutual friends are connected)

b) User Community ($U$)
( ─────── Normal link,
- - - - - - - Suspicious Link)

**Figure 4:** Mutual Clustering Coefficient ($M_{cc}$) on a sample graph.

Here, $CC_v$ is the clustering coefficient of a user $v$, $k_v$ is the number of friends of $v$ and $l_v$ is the number of edges between the $k_v$ neighbors/friends of $v$. However, in our case, we have modified the clustering coefficient metric to compute the extent to which mutual friends of two users are themselves connected rather than the extent to which only neighbors of a standalone user are connected, and we call this as Mutual Clustering Coefficient ($M_{CC}$). In number of edges between the $m_{u,v}$ mutual friends of $u$ and $v$.

The $M_{cc}$ values range from 0 to 1, where 0 indicates that none of the mutual friends of two corresponding users are connected and 1 indicates all the mutual friends of two friends are connected to each other. The higher is the value of $M_{cc}$, the strongly are the mutual friends connected. In order to better understand the Modified clustering coefficient computation, we have presented a small example in the Figure 4. The table 4 (a) shown in the figure holds the $M_{cc}$ values calculated from a conceptual user community shown in 4 (b) using the equation (2). In the given sample graph as shown Figure 4 (b), the dotted lines represent the suspicious links established by fake profiles and normal links between legitimate users are represented by solid lines. The complete pseudo-code for calculating the mutual clustering coefficient ($M_{cc}$) for two connected users on a social network is shown in Figure 5.

---

$Algorithm\ 2 - Mutual\_Clustering\_Coefficient\ (M_{CC})$
{
  **Input** : $friend\_mat$
  $ltm = lower\_traingular\_matrix(friend\_mat)$     /* for bidirectional social networks only
  $links = [\ ][2]$     /* holds edge information of a user pair
  for $(i = 1\ to\ nrows(ltm))$
    for $(j = i\ to\ ncols(ltm))$
      if $(ltm[i,j] == 1)$
        add $(i,j)$ to link
  $m_{cc}list = [\ ][3]$     // holds mutual clustering coefficient of connected user pairs
  for $(i = 1\ to\ nrows(links))$ {
    $u = links[i][1]$     /* vertex of edge
    $v = links[i][2]$     /* vertex of edge
    $mfi = [\ ]\ \ mfi = indicesOfMutualFriends(u,v)$  /* returns common friends of a connected user pair
    $m_{uv} = length(mfi)$   /* returns the number of of mutual friends
    $l_{uv} = 0$     /* $l_{uv}$ holds the number of edges between the $m_{u,v}$ mutual friends of u and v
    if $(m_{uv} > 1)$
      for $(i = 1\ to\ (m_{uv} - 1))$
        for $(j = i + 1\ to\ m_{uv})$
          if $(friend\_mat[mfi[i], mfi[j] == 1)$
            $l_{uv} = l_{uv} + 1$
    $mcc_{uv} = (2 * l_{uv})/\ m_{uv}\ *\ (m_{uv} - 1)$ /* calculating mutual clustering
    Add $(u, v, mcc_{uv})$ to $m_{cc}$ list
  }
}

---

Figure 5: Pseudo-code to calculate the $M_{CC}$ for every connected user pair in a community

Unlike Twitter network, the friendship in the Facebook is bidirectional in nature therefore, we are calculating the lower triangular matrix ($ltm$) from the input matrix otherwise the same edge will be counted twice. The $links$ array holds the information related to edges, i.e. the nodes (profiles) which have established the edge between them. $Mcc_{uv}$ holds the mutual clustering coefficient value for users $u$ and $v$ and $m_{cc}list$ sotres $M_{cc}$ values for all connected user pairs in the community.

The function $indicesOfMutualFriends(u,v)$ calculates the friend list of two users (say u and v) and then returns the indices of common friends between these two users, function $m_{uv}$ hold the total number of mutual friends between two connected users and $l_{uv}$ holds the number of links within the mutual friends of $u$ and $v$. In order to calculate the $M_{CC}$ for every connected user pair within a community, we need a matrix (*friend_mat*) which represents the network structure of every connected pair in the community. This matrix is supplied as an input to the $M_{cc}$ algorithm to yield the mutual clustering coefficient value for each pair of connected users.

The fourth column of the feature vector stores the Profile Similarity Score ($PSS$) between each pair of connected users in the community.

The similarity score can be computed based on varied profile attributes of a user such as $Work$ ($w$), Education ($e$), Hometown ($ht$), current_city ($cc$), etc.

For calculating the similarity between two profiles, by considering any possible re-organization of the structure of words in the extracted features, one common way is to employ fuzzy similarity measure. For constructing features for the proposed work, we have used Fuzzywuzzy [25], a python library which contains a number of functions for fuzzy string matching. Basically, Fuzzy string matching finds strings that approximately match a given pattern. It makes use of Levenshtein Distance [26] to calculate the similarity between sequences of words. The fuzzywuzzy library contains several functions for different applications of String Matching. The value returned by each function ranges between 0 to 100 with 0 indicating totally different sequences and 100 represents exact similarity. We applied several functions and conclude to the one where we observed best results as per the description of the sequences. The following figure shows some of the functions in the fuzzywuzzy library which we have tested on our data.

```
"******************** Case 1 ********************"
from fuzzywuzzy import fuzz

fuzz.ratio('Research Laboratory, India','Research Laboratory, Social Networks, India')
75
fuzz.partial_ratio('Research Laboratory, India','Research Laboratory, Social Networks, India')
88
fuzz.token_sort_ratio('Research Laboratory, India','Research Laboratory, Social Networks, India')
76
fuzz.token_set_ratio('Research Laboratory, India','Research Laboratory, Social Networks, India')
100
"******** Case 2 ***************"
fuzz.ratio('Research Laboratory, Bangalore, India','Research Laboratory, Online Social Networks, Bangalore, India')
76
fuzz.partial_ratio('Research Laboratory, Bangalore, India','Research Laboratory, Online Social Networks, Bangalore, India')
73
fuzz.token_sort_ratio('Research Laboratory, Bangalore, India','Research Laboratory, Online Social Networks, Bangalore, India')
75
```

```
fuzz.token_set_ratio('Research Laboratory, Bangalore, India','Research Laboratory, Online
Social Networks, Bangalore, India')
100
```

As shown in code snippet written in Python programming language, we have tested four string matching functions for two random strings from our dataset and out of all the four functions the fuzz.token_set_ratio ( ) function gave the best result in both the test cases (Case1 and Case 2).

The token functions divide the string based on white-spaces, change all uppercase letters into lower-cases and remove the stop words (non-alpha and non-numeric characters). These functions tokenize the string and treat it as a set or a sequence of words. It should be noted here that we have already applied these text processing techniques in the data preprocessing section because of several reasons, first, it will be easy for the string matching functions to process if an appropriate input is provided to them. Second, it decreases the processing time for similarity calculation**.**

For our work, we chose the fuzzy.token_set_ratio as the function to determine the similarity between two profiles as it performs the similarity calculation better these fake profiles are suspicious and based on this we created one more column to the data matrix namely class which labels the user-links as 0 (for real) and 1 (for suspicious).

Finally, we prepared a training/testing dataset with five columns as shown in Table 2. The first two columns hold the information about the connected user pair, the third and fourth column holds the $M_{CC}$ and $PSS$ values respectively for every connected user pair in the community. Note that the $PSS$ holds the four attributes ( $Work\ (w)$, Education $(e)$, Hometown $(ht)$, $Current\_city(cs)$ ) and value of each attribute ranges from 0 to 1. And the last column holds the label for the corresponding connection with 1 indicating normal link and 0 represents the suspicious connection.

| $Node\ (u)$ | $Node\ (v)$ | $M_{CC}$ | $PSS$ | | | | $Class$ |
|---|---|---|---|---|---|---|---|
| | | | $wk$ | $ed$ | $ht$ | $cs$ | |
| $User\ (u)$ | $User\ (v)$ | $0-1$ | $0-1$ | $0-1$ | $0-1$ | $0-1$ | $0\ or\ 1$ |

**Table2:** Logical representation of the data-matrix to be supplied to the classifier (Profile Similarity Score ($PSS$) based on four attributes ($Work\ (w)$, Education $(e)$, Hometown $(ht)$, $Current\_city(cs)$) ) out of all the other functions. In our collected data set the users have provided multiple values for different fields, for example, in the *education* attribute, some of the users have provided details about their schooling, college, post-graduation, and Ph.D. while others have mentioned about *current education* only. The value of fuzzy.token_set_ratio() the function is not affected (decreased) by random words that might exist inside the strings and therefore proves to be best for various kinds of analysis. Furthermore, it evades cases of typical obfuscation techniques as it searches for similar sets of terms that refer to the entity.

     Since we have injected fake profiles into the targeted user community, all the links connected to

### 3.3.1 Feature Analysis

In order to visually analyze the distinguishing power of features for the suspicious link detection process, all the features have been shown in figure 6. The figure 6(a) depicts the mutual clustering coefficient $M_{cc}$ which measures the connectivity of mutual friends between a friend pair. Since we are focusing on the detection of links established by adversaries by exploiting the mutual friend feature, therefore, value for $M_{CC}$ will remain higher for suspicious connections. The most of the suspicious connections have been observed with $M_{CC}$ higher than 0.45 whereas the normal links have $M_{CC}$ values higher than 0.15. We observed that the value of $M_{cc}$ for suspicious links is higher than the average of value $M_{CC}$ for normal links. We recorded the average $M_{cc}$ value as 0.4377 and 0.5521 for normal and suspicious connections respectively. Based on our hypotheses ("if the $M_{cc}$ for any links is high along with low profile similarity score, the link is considered as suspicious"), we visualized other profile features. As we can see from figure 6(b) and 6(c) the work and education similarities between normal (real) links are mostly high unlike suspicions ones, since most of the connections are being established on a network if the two persons know each other in offline (either at some educational institute or work organization, in our case), which is not true between the real and fake users. As observed, the suspicious link users have work and educational similarity not more than 0.45 and 0.25 respectively.

As we have collected the data from a group on Facebook network which contains faculty members and research scholars, therefore, there are chances that they have different educational backgrounds and are mostly working for the same organization, and this is the reason that unlike figure 6(b), the figure 6(c) shows several instances at bottom (Level 0.0).

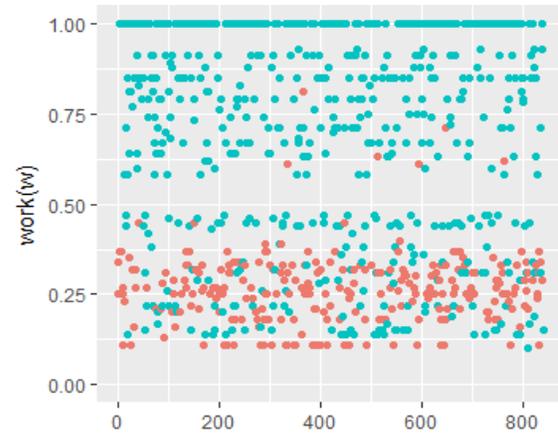

$b)\ Work\ (w)$

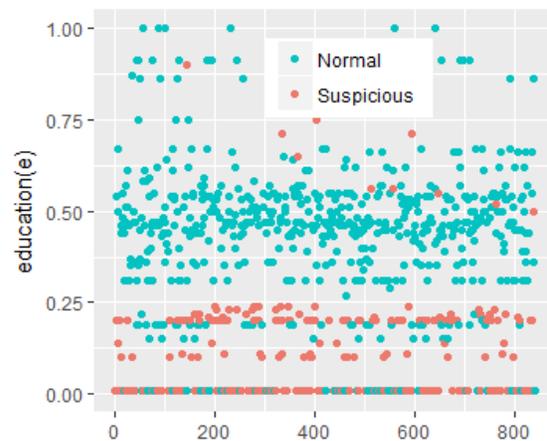

$c)\ Education\ (e)$

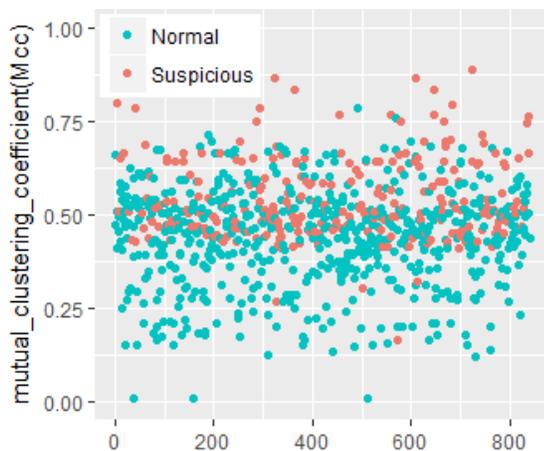

$a)\ Mutual\ Clustering\ Coefficient\ (M_{CC})$

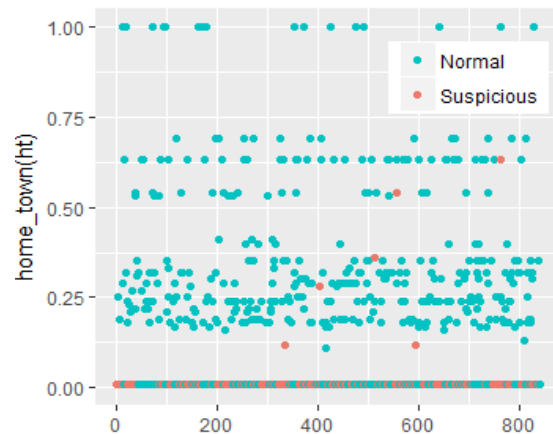

$d)\ Home\ Town\ (ht)$

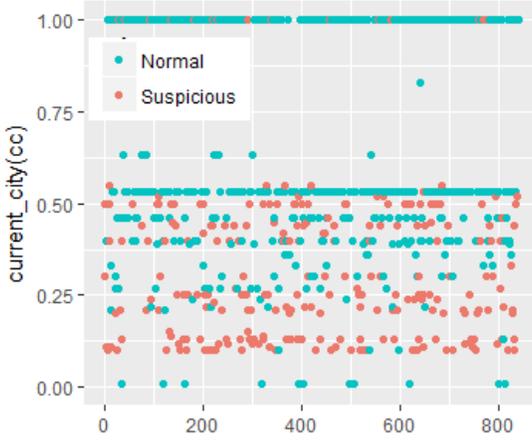

$e)\ Current\ City\ (cc)$

**Figur 6:** Statistical Analysis of features for both real and suspicious links shown in teal and orange colors respectively. The similarity score for every feature is shown along the $y-axis$ and the numbers of links are plotted along $x-axis$

Furthermore, it is clearly visible in figure 4(b), that the similarity values for suspicious links for $work(w)$ attribute are lower than the average of similarity values for normal links.

The average similarity score based on work(w) attribute for normal and suspicious links has been observed as 0.7143 and 0.2679 respectively, whereas for $education(e)$ attribute the average similarity score is 0.4323 and 0.1111 for normal and suspicious links respectively. The figure 6(d) shows the home_town similarity where most of the suspicious links rest at the bottom (0.0 level), whereas figure 6 (e) shows the similarity based on $current\_city(cc)$ is lower than 0.5 for suspicious links. Furthermore, for normal links both the $home\_town(ht)$ and $current\_city(cc)$ based similarity scores range from 0.0 to 1.0. Also, it can be noticed from the figure 6(e), the $current\_city(cc)$-based similarity values for suspicious links are lower than the average of similarity values for normal links. The average similarity score for $home\_town\ (ht)$ attribute has been recorded as 0.2223 for normal links and 0.0081 for suspicious links, whereas the average similarity score for $current\_city(cc)$ attribute is 0.3558 and 0.6972 for suspicious and normal links respectively. Furthermore, it can be clearly observed from figure 6(b) to 6(e), the normal (positive) links have highest attribute similarity than suspicious ones, whereas the $M_{cc}$ value is higher for suspicious links than normal connections.

### 3.4 Suspicious Link Classification System

The prepared data set from the previous subsection is used for training of suspicious link classifier. One-third portion of the prepared data set has been separated for testing the accuracy of the proposed system. Remaining two-third portion was used to train and select the model following 10-fold cross-validation. A Number of classification techniques including Decision Tree, Support Vector Machine (SVM) and Naïve Bayes (NB) have been used for designing classifier.

The decision tree algorithm, J48, classifies the instances by building a decision tree on the bases of the training data. The root node of the tree is the feature which has the highest Information Gain. The leaf of the tree describes the decision which is dependent on other independent nodes of the tree. The decision tree generated for our collected dataset is shown in Figure 7 and it is clear from the figure that all the features f1-f5 contributed to the decision making process. Feature f3 is the root node of the tree as it has the highest classification power.

Support Vector Machine (SVM) is a discriminative classifier which on given labeled training data produces an optimal hyperplane and classifies the new instances. The hyperplane is a line that linearly separates and classifies a set of data. The distance between the hyperplane and the nearest data point from either class is known as the margin. In order to enhance the chances of new instances being classified correctly, the goal is to find a hyperplane with the greatest possible margin between and any training data point and hyperplane.

We have employed a non-linear support vector machine (SVM) classifier [35] with the Radial Basis Function (RBF) kernel using package e1071

[37] on different values of gamma and $'c'$ which controls the degree of nonlinearity and over-fitting of the model respectively, with the help of grid search policy to find the highest classification accuracy.

($M_{cc}$, w, e, ht, cc). As the P(X) is a normalizing factor and is equal for both the classes therefore, in order to perform the classification we need to maximize the numerator only.

```
f3 <= 0.24
|   f2 <= 0.68
|   |   f5 <= 0.63
|   |   |   f5 <= 0.36: Suspicious (137.0/1.0)
|   |   |   f5 > 0.36
|   |   |   |   f2 <= 0.37
|   |   |   |   |   f2 <= 0.17
|   |   |   |   |   |   f2 <= 0.14: Suspicious (10.0)
|   |   |   |   |   |   f2 > 0.14: Normal (6.0/1.0)
|   |   |   |   |   f2 > 0.17: Suspicious (71.0)
|   |   |   |   f2 > 0.37
|   |   |   |   |   f2 <= 0.38: Normal (6.0)
|   |   |   |   |   f2 > 0.38: Suspicious (2.0)
|   |   f5 > 0.63
|   |   |   f2 <= 0.21: Normal (38.0/2.0)
|   |   |   f2 > 0.21
|   |   |   |   f3 <= 0.11
|   |   |   |   |   f1 <= 0.508333: Normal (3.0/1.0)
|   |   |   |   |   f1 > 0.508333: Suspicious (18.0)
|   |   |   |   f3 > 0.11: Normal (3.0)
|   f2 > 0.68: Normal (46.0)
f3 > 0.24
|   f3 <= 0.49: Normal (265.0)
|   f3 > 0.49
|   |   f5 <= 0.52
|   |   |   f2 <= 0.62: Suspicious (7.0/1.0)
|   |   |   f2 > 0.62: Normal (47.0/2.0)
|   |   f5 > 0.52: Normal (180.0/3.0)
```

$f1: M_{cc}$

$f2: Work(w)$

$f3: Education(e)$

$f4: Home\_Town(ht)$

$f5: Current\ City(cc)$

Figure 7: Decision Tree generated from collected dataset.

Naive Bayes, based on Bayes theorem is a simple probabilistic classification approach which assumes that every feature is independent of the values of other features and this assumption is called class conditional independence. For the classification of instances, the posterior probability is computed for each class [34] as

$$P(Y|X) = \frac{P(X|Y)P(Y)}{P(X)} \quad (3)$$

Where $Y$ is the class variable and $X$ is the feature vector. In our case, the class label holds two classes that is $Y = ("normal", "suspicious")$ and $X$ holds

Once the training is complete the performance of the trained model is evaluated using testing dataset. The commonly used performance evaluation measures for classification models were employed such as Precision, Recall, Accuracy, F-measure, Receiver Operating Characteristic (ROC) curve, etc.

|  |  | Actual | |
|---|---|---|---|
|  |  | Normal | Suspicious |
| **Predicted** | Normal | a | b |
|  | Suspicious | c | d |

Table 3 Confusion matrix

All these measures have been briefly discussed in this section in order to describe the goodness of trained model. In the evaluation we considered the confusion matrix shown in Table 3 where $a$ represents the total number of normal links correctly classified by the model, $b$ refers the number of normal links misclassified, $c$ refers the number of suspicious links incorrectly classified and $d$ refers the number of suspicions links correctly classified. Based on this confusion matrix we have calculated the precision, recall and F-measure and accuracy in order to evaluate the performance of the proposed model. Precision (P) is the ratio of a number of instances correctly classified to the total number of instances. By using the confusing matrix in table 3 it can be calculated as follows:

$$Precision\ (P) = \frac{d}{c+d} \quad (4)$$

Recall (R) is the ratio of a number of instances correctly classified to the total number of predicted instances. Recall can be calculated using the formula as

$$Recall\ (R) = \frac{d}{b+d} \quad (5)$$

Accuracy refers to the ratio of number of instances correctly predicted to the total of instances predicted by the model as is denoted as

$$Accuracy = \frac{a+d}{(a+b+c+d)} \quad (6)$$

F-measure is the harmonic mean between precision and recall and can be calculated as follows

$$F-measure = \frac{2PR}{(P+R)} \quad (7)$$

Since F-measure is a value that summarizes both the precision and recall therefore it is considered as more supportive in evaluating the classifier's performance than the former two measures.

The receiver operating characteristic (ROC) curve is used to evaluate the diagnostic performance of the classifier. The curve is created by plotting the true positive rate (TPR) against the false positive rate (FPR) at various threshold settings. The TPR is also known as recall or probability of detection whereas the FPR is also known as the probability of false alarm, calculated as under

$$False\ Positive\ Rate\ (FPR) = \frac{b}{a+b} \quad (8)$$

Out of the three selected classifiers, J48 achieved the highest results with 99.6% of accuracy. Table 4 illustrates the confusion matrix obtained by J48 classifier with 99.4% of normal links and 100% suspicious links correctly classified, leaving a very small percentage of normal links misclassified. Table 5 holds the precision, recall, and F-measure for both the normal and suspicious links calculated from the confusion matrix generated by the J48 algorithm.

|  |  | Actual | |
|---|---|---|---|
|  |  | Normal | Suspicious |
| Predicted | Normal | 99.4% | 0.6% |
|  | Suspicious | 0% | 100% |

Table 4 Confusion matrix generated by J48

|  | Precision | Recall | F-measure |
|---|---|---|---|
| Normal Link | 0.99 | 1 | 0.99 |
| Suspicious Link | 1 | 0.99 | 0.99 |

Table 5 Evaluation Matrices for J48 algorithm

Furthermore, we have also compared the result obtained by J48 with two more classifiers: Support Vector Machine (SVM) and Naïve Bayes (NB) with the help of implementation provided by R.

The evaluation matrices including precision, recall, F-measure, and accuracy have been calculated for each classifier for suspicious as well as normal category as illustrated in Table 6. It is clearly visible in the table that other two classifiers have also achieved better accuracy as 98.01% and 93.63% by SVM and Naïve Bayes classifier respectively. This proves that our proposed features have enough ability to distinguish the suspicious links from normal ones on Facebook network. For SVM-based classification, we have used SVM with Radial Based Function

(RBF) kernel on different $c$ and $gamma$ values and finally achieved the F-measure with 0.98 and 0.96 for normal and suspicious links respectively with 98% accuracy at $sigma = 5$ and $c = 9$.

accuracy for the evaluating and comparing classifiers [38].

|  | Precision | | Recall | | F-measure | | Accuracy |
|---|---|---|---|---|---|---|---|
|  | **Normal** | **Suspicious** | **Normal** | **Suspicious** | **Normal** | **Suspicious** |  |
| Decision Tree (J48) | **0.99** | **1** | **1** | **0.99** | **0.99** | **0.99** | **99.60** |
| Support Vector Machine (SVM) | 0.98 | 0.97 | 0.99 | 0.96 | 0.98 | 0.96 | 98.01 |
| Naïve Bayes (NB) | 0.99 | 0.83 | 0.91 | 0.99 | 0.95 | 0.90 | 91.63 |

**Table 6** Evaluation matrices of three Classification Techniques

The Naïve base classifier generates the results slightly poor than other two techniques it may be because of its naive assumptions that all the variables are uncorrelated to each other but in our case, it is not true, features are dependent on each other.

A performance comparison of multiple classifiers has been given by plotting ROC against test data set as shown in Figure 8.

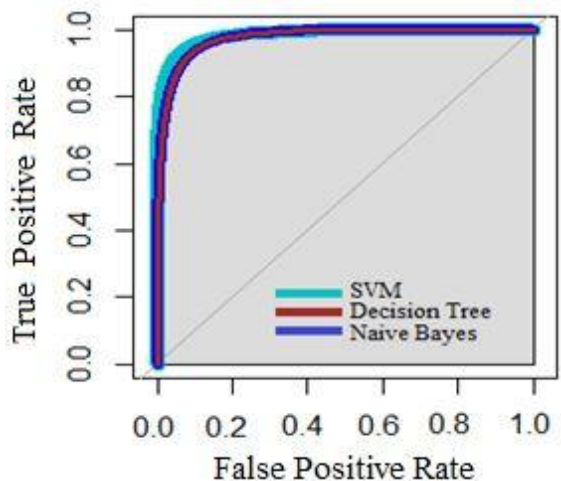

**Figure 8:** ROC for test dataset

Moreover, Figure 8 calculated the AUROC (Area under ROC) as shown in Table 7, to know statistic for the goodness of proposed classifier. AUROC has been observed as a better performance measure than

| Algorithm | AUROC | Accuracy |
|---|---|---|
| Decision Tree (J48) | 0.9985 | 99.6 |
| Support Vector Machine (RBF) | 0.9830 | 98.01 |
| Naïve Bayes | 0.9813 | 91.63 |

**Table 7:** AUROC obtained for different classifiers

If the prediction is totally random, AUC would be 0.5 which is considered as below than poor.

As per the experimental results are shown in table 7, the decision tree with AUC value 0.99 performs the best based on both metrics. The nonlinear classifier, SVM with RBF kernel and Naïve base also performances well in terms of its AUC value. It is to be noted that the SVM and Naïve Bayes shows different performance in terms of accuracy but based on AUC measure both the algorithms are identical.

## 4. Conclusion

In this paper, we have proposed an approach for the detection of the suspicious link based on number of features which are combination of Mutual Clustering Coefficient ($M_{cc}$) and profile information of a user. The experiments demonstrate that the proposed features are conducive to detect user connections (links) which have been established by fake users within a user community on the Facebook network by exploiting the mutual friends feature. We have formulated and introduced a unique and a novel feature: the $Mutual\ Clustering\ Coefficient\ (M_{CC})$, by modifying clustering coefficient concept to analyze the extent to which the common neighbors (mutual friends) of two friends are connected to each other.

Furthermore, different similarity measures have been applied on varied profile features to compute the profile similarity score ($PSS$) of each user pair in the collected dataset. The $M_{CC}$ along with four profile features viz ($work(w),\ education(e),\ hometown(ht),\ current\ city(cc)$) have proved to be optimal features for the detection of suspicious links on Facebook network as all the three classifiers ($Decision\ Tree, SVM\ and\ Naive\ Bayes$) have shown good results on test dataset.

The proposed model can be used by the OSN service providers to alert their members with a list of suspicious connection (links) from their respective friend lists so that users can themselves verify the suggested links and filter their friend list as per their requirement. Although, the proposed approach has been tested for the Facebook users, with the little modifications it will be applicable to other social network sites as well. One of the future works will be to extend this system for the Twitter and LinkedIn like networks, where fake links may be more vulnerable. Furthermore, we have collected data from 1000 interconnected Facebook profiles for designing the proposed classifier, one of the future extensions will be to further strengthen the count of profiles in the dataset and make it publicly available to other researchers for their study.